\documentclass[letter]{aa} 

\usepackage{graphicx}
\usepackage{float}
\usepackage{txfonts}
\usepackage{natbib}
\usepackage{amsmath}
\usepackage{amsfonts}
\usepackage{amssymb}
\usepackage{graphicx}
\usepackage{lmodern}
\usepackage{subcaption}
\usepackage{siunitx}

\newcommand{\cbn}[1]{C$_{#1}$}  

\bibpunct{(}{)}{;}{a}{}{,}

\begin{document}

\title{Mapping the structural diversity of C$_{\sf 60}$ carbon clusters and their infrared  spectra
}
\titlerunning{IR carbonaceous clusters}

\author{C. Dubosq,\inst{1} C. Falvo,\inst{2,3} F. Calvo,\inst{3} M. Rapacioli,\inst{1} P. Parneix,\inst{2}  T. Pino,\inst{2} 
 A. Simon$^{\star}$\inst{1}  }
\authorrunning{Dubosq et al.}

\offprints{\\ \email{aude.simon@irsamc.ups-tlse.fr}}
\institute{Laboratoire de Chimie et Physique Quantiques LCPQ/IRSAMC, Universit\'e de Toulouse and CNRS, UT3-Paul Sabatier,  118 Route de Narbonne, F-31062 Toulouse, France 
\and
Institut des Sciences Mol\'eculaires d’Orsay (ISMO), CNRS, Univ. Paris Sud, Universit\'e Paris-Saclay, 91405 Orsay, France
\and
Univ. Grenoble Alpes, CNRS, LIPhy, 38000 Grenoble, France}

\date{Received ?; accepted ?}

\abstract 
{The current debate about the nature of the carbonaceous material carrying the infrared (IR) emission spectra of planetary and proto-planetary nebulae, including the broad plateaus, calls for further studies on the interplay between structure and spectroscopy of carbon-based compounds of astrophysical interest. The recent observation of C$_{60}$ buckminsterfullerene in space suggests that carbon clusters of similar size may also be relevant. In the present work, broad statistical samples of C$_{60}$ isomers were computationally determined without any bias using a reactive force field, their IR spectra being subsequently obtained following local optimization with the density-functional-based tight-binding theory. Structural analysis reveals four main structural families identified as cages, planar polycyclic aromatics, pretzels, and branched. Comparison with available astronomical spectra indicates that only the cage family could contribute to the plateau observed in the 6--9 \si{\micro\meter} region. The present framework shows great promise to explore and relate structural and spectroscopic features in more diverse and possibly hydrogenated carbonaceous compounds, in relation with astronomical observations.}

\maketitle

\section{Introduction} \label{int}
Buckminsterfullerene (\cbn{60}) has  recently been identified in space \citep{cami2010,sellgren10}. Although recent observational studies  suggest that fullerenes could be produced by photochemical dehydrogenation and isomerization of amorphous carbon grains \citep{Hernandez2010,Hernandez2011,Hernandez2011b,Hernandez2012}, the composition of these grains is still debated. Infrared (IR) spectroscopic observations of fullerene-rich planetary nebulae (PNe) found a broad plateau with substructure at 6--9 \si{\micro\meter} range \citep{Hernandez2010,Hernandez2011,Hernandez2011b,Hernandez2012,Salas_ApJ2012}. The origin of this plateau remains elusive and various hypotheses have been formulated about its possible carriers, including hydrogenated amorphous carbon (HAC) compounds \citep{Salas_ApJ2012}, very small grains, or polycyclic aromatic hydrocarbon (PAH) clusters \citep{tielensrev,Buss2013,Rapacioli2005}. Similarly, the  8 and 12 \si{\micro\meter} plateau features of some proto-planetary nebulae (PPNe) were assigned to the vibrations of alkanes or alkyle side groups pertaining to large carbonaceous particles \citep{Kwok2001,Hrivnak2000}, whose details also remain unclear.

These observations and their various interpretations have motivated further studies to unravel the molecular nature of the carriers of these spectral features. The present letter describes a purely computational framework designed to systematically and extensively explore the structural diversity of \cbn{60} isomers in a statistically unbiased way and to relate the various isomers obtained to their IR spectral features. Sorting these isomers using appropriate order parameters, our mapping provides unprecedented correlations between structure and spectroscopy in astrochemically relevant compounds.
This study also brings indirect insight into the possible precursors of stable fullerenes under astrophysical environments, whose formation in circumstellar environments of carbon-rich evolved stars is somewhat puzzling \citep{Salas_ApJ2012}. Two main routes towards such highly organized molecules have been proposed in the literature, consisting of either the successive growth from smaller carbon chain building blocks, known as the closed network growth mechanism \citep{Nat_Comm_2012_CNGfull}, or alternatively the decay and rearrangement of larger carbon grains or dehydrogenated PAHs \citep{Berne2015} under the action of ultraviolet photons \citep{Zhen_2014} or by collisions with energetic particles  \citep{Omont_rev}. These so-called bottom-up and top-down mechanisms are  expected to prevail in the hot,  dense envelopes of evolved stars and in low-density environments such as interstellar clouds, respectively \citep{Berne_Tielens2012}.

The  letter is organized as follows. In Section \ref{met}, we briefly describe the computational approach. In Section \ref{res} we analyze the structural diversity of \cbn{60}, the classification into four main families of isomers, and the relation with IR features. In Section \ref{disc} we discuss the possible contribution of these families to some features of astronomical spectra for selected PNe and PPNe. Finally, we make some concluding remarks  in Section \ref{con}.


\section{Computational details} \label{met}

 Following the simulation protocol recently described  in~\citet{bonnin2019}, a systematic and unbiased search for isomers of \cbn{60} was computationally undertaken using the efficient but realistic reactive empirical bond order (REBO)  force field \citep{Brenner:2002xy}. Replica-exchange molecular dynamics (REMD) simulations were performed in the 2500--6500~K temperature range using 16 trajectories, the temperatures being allocated in a 12-member geometric series with four extra temperatures
around the C$_{60}$ melting temperature~\citep{Pretzel_prl1994}. All replicas were initiated from the buckminsterfullerene global minimum. The use of high temperatures and the REMD framework ensures that the sample is not biased towards this starting point. The trajectories were integrated over 100~ns using a time step of 0.1~fs, and configurations were periodically saved and locally reoptimized. Spherical boundary conditions were used and the density was varied over four different simulations with $\rho=0.025,0.15,0.4,$ and 1.7~g$.$cm$^{-3}$. Dissociated structures corresponding to a threshold distance greater than 1.85~\AA\ were excluded. This unbiased sampling stage produced 656\,438 nonequivalent isomers.

While REBO provides a reasonable description of intramolecular forces in carbon nanostructures, it lacks electrostatics and cannot provide vibrational spectra. Instead we used the self-consistent charge (SCC) density-functional-based tight-binding (DFTB) method employing the mio set of parameters \citep{elstner98} with additional dispersion corrections \citep{DFTB_CM3}. Although significantly more expensive than REBO, the SCC-DFTB approach remains computationally efficient owing to approximations and pre-computed pair integrals. It was previously employed successfully to determine structural \citep{CPL2015_Irle, JCP2015_DFTB_CClusters, JCP2006_Irle, JPCA2007_Irle},
spectroscopic \citep{joalland2010,Simon_matrix2017,SimonPCCP2012,SimonJCP2013,SimonCOMP2013}, and fragmentation \citep{Simon20160195,Simon2017_dissoc,PAH+H} properties of PAH molecules and oligomers of astrophysical interest under various charge and energy states, at and away from equilibrium. The formation of PAHs or carbon clusters \citep{ACSNano2009_Irle,JCP2010_Irle} and their hydrogen uptake capability \citep{Carbon2018_Irle} were also scrutinized.

Reoptimization of the REBO isomers produced 309\,168 independent structures at the SCC-DFTB level of theory, for which the IR absorption spectra were computed in the double harmonic approximation. The raw vibrational frequencies obtained with SCC-DFTB were scaled 
to match the best reference data (see Appendix A). Owing to this dependence of the scaling factors on the frequency domain, a discontinuity arises near  6.0~\si{\micro\meter}, preventing  an accurate analysis of the  [5.8-6.2 $\mu$m] domain.

All REBO simulations were performed using the LAMMPS software package \citep{Plimpton:1995aa}; the SCC-DFTB calculations were carried out with the deMonNano software package \citep{demonnano}.

\section{Structural partitioning} \label{res}

The structural properties of our sample of isomers were revealed by employing a reduced number of parameters that characterize features at different scales. In addition to the SCC-DFTB energy, we considered the shape of the individual isomers through the radius of gyration $r_g$ and the Hill-Wheeler asphericity parameter $\beta$ \citep{Hill_Wheeller_PRB1853}. The radius of gyration $r_g$ measures the spatial extension of the isomer, while $\beta<1$ measures deviations to the perfect sphere for which $\beta=0$. As is shown below, the Hill-Wheeler parameters, previously used to characterize the shape of metallic clusters \citep{Calvo_PRB2000}, appear to be quite sensitive for carbon compounds as well.

Local ordering in the carbon clusters was examined from their hybridization level. From the SCC-DFTB electronic density matrix, Mayer's bond orders \citep{Mayer_CPL1983} were computed and bonds placed accordingly when the bond order was higher than 0.8. A carbon atom with a coordination number of $n+1$ is attributed to a sp$^{n}$ hybridization level. For a given isomer, the fraction of sp$^2$ atoms averaged over all atoms was assigned as the main local order parameter to measure the extent of sp$^2$ hybridization.

A last discriminatory parameter, which can be seen as a semi-local or semi-global index, is the number of 6-member rings within each isomer. This quantity is related to the extent of connectivity within the structure, and can also differentiate isomers with large aromatic domains from those exhibiting topological defects.
\begin{figure*}
\centering
\includegraphics{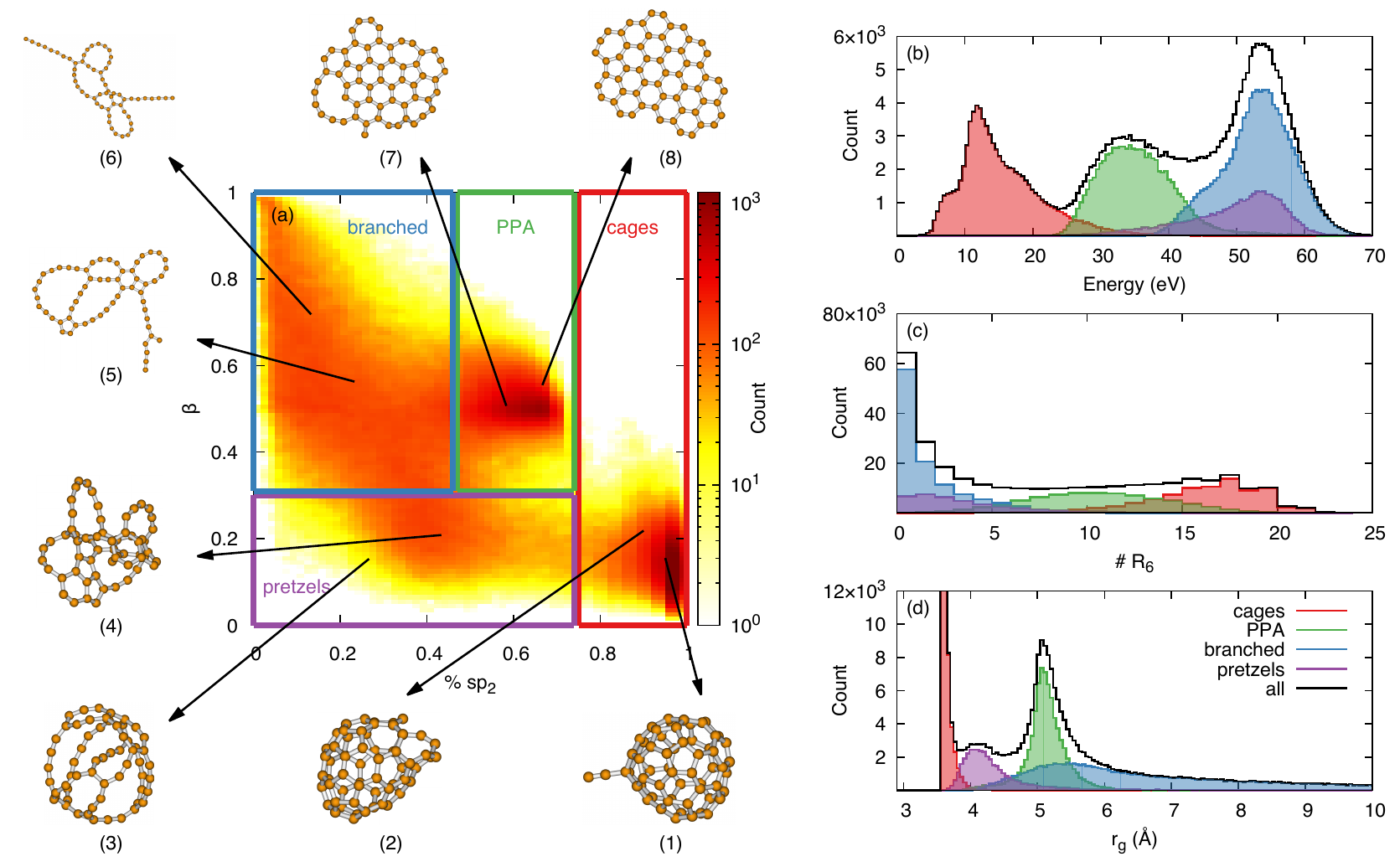}
\vspace*{-0.3cm}
\caption{Distributions of the samples of isomers based on specific order parameters. (a) Two-dimensional distribution as a function of the sp$^2$ fraction and asphericity $\beta$. Panels (b), (c), and (d) show the one-dimensional distributions as a function of isomer energy, number of 6-member cycles, and gyration radius, respectively. In panel (a) the boxes classify the four structural families, cage, planar polycyclic aromatic (PPA), pretzel, and branched;   the same colors are used in panels (b-d). Two  example structures (numbered 1-8) are shown for each family. \label{fig1}}
\end{figure*}
Figure \ref{fig1} shows how the various order parameters are distributed among the  $3\times 10^5$ isomers. Among the five parameters considered, the fraction of sp$^2$ atoms and the asphericity $\beta$ are found to best classify the various isomers into  structural families. Four such families are identified, all fitting into rectangular boxes in Fig.~\ref{fig1}(a). Each family has generally clear signatures as well-defined peaks in the one-dimensional distributions of Figs.~\ref{fig1}(b-d) as a function of energy, number of 6-member cycles, and gyration radius, respectively.

The first family, which is  thermodynamically the most stable (lowest energies), is defined by isomers with the highest fraction of sp$^2$ atoms (above 75\%). They  generally show a high spherical character (low $\beta$). These isomers also have the highest number of 6-member rings but the lowest gyration radius (3.7~\AA), indicating a rather compact shape. These 8.2$\times$10$^4$ isomers, known as cages, include buckminsterfullerene as their most remarkable member, as well as defective fullerenes such as  structure (2) in Fig.~\ref{fig1} (a).

In order of increasing energy, the second family consists of intermediate fractions of sp$^2$ atoms  between 45\% and 75\%, and an asphericity parameter $\beta$ higher than 0.3. These isomers exhibit clear peaks in the energy and gyration radius distributions, and once resolved they also show intermediate values in the number of 6-member rings. This family, classified as flakes or PPAs, contains 7.7$\times$10$^4$ members. 

One minor peak in the gyration radius distribution near 4~\AA\ points at a specific family with rather compact structures, correlated with low asphericity but a more arbitrary sp$^2$ fraction (still lower than 75\%). This third family is defined in the two-dimensional distribution from $\beta<0.3$, and includes rather spherical but fairly open structures with many carbon chains. Following \citet{Pretzel_prl1994} we classify the 3.8$\times$10$^4$ members of this family as pretzels.

Finally, all the remaining isomers lie in the range of $\beta>0.3$ and with a sp$^2$ fraction lower than 45\%. They are associated with the highest energies, the lowest number of 6-member rings, but the highest gyration radius, whose distribution is particularly broad. They usually contain multiple linear chains connected at a limited subset of atoms. This family is broadly referred to as made of branched structures, which are found in numbers of 1.1$\times$10$^5$.

Selected isomers are depicted in Fig.~\ref{fig1} to further illustrate the terminology employed above. For each family, the statistically averaged values of all  five energetic or structural parameters was determined, and the Euclidean distance of each member of the family to this five-dimensional point calculated. The member having the lowest distance was identified as a most representative isomer of the family, without any bias from human judgment. 
The structures numbered (1), (4), (5), and (7) in Fig.~\ref{fig1} are the statistically most representative isomers. For each family another isomer was chosen to further illustrate the structural diversity within the families. These isomers are numbered (2), (3), (6), and (8).

The structural diversity found by exploring the potential energy surface of C$_{60}$ suggests an equally diverse set of rearrangement pathways towards fullerenes. In addition to  amorphous \citep{Sinitsa2017} and PPA \citep{Berne2015} intermediates, linear carbon chains, as already identified in circumstellar environments \citep{Bernath_1989,Hinkle1988}, could also play a role by forming pretzel or branched structures also identified here at higher energies. In view of their possible spectroscopic detection, it is important to now determine their infrared signature.

\section{IR spectra in comparison with astronomical data} \label{disc}

\begin{figure}
\centering
  \includegraphics{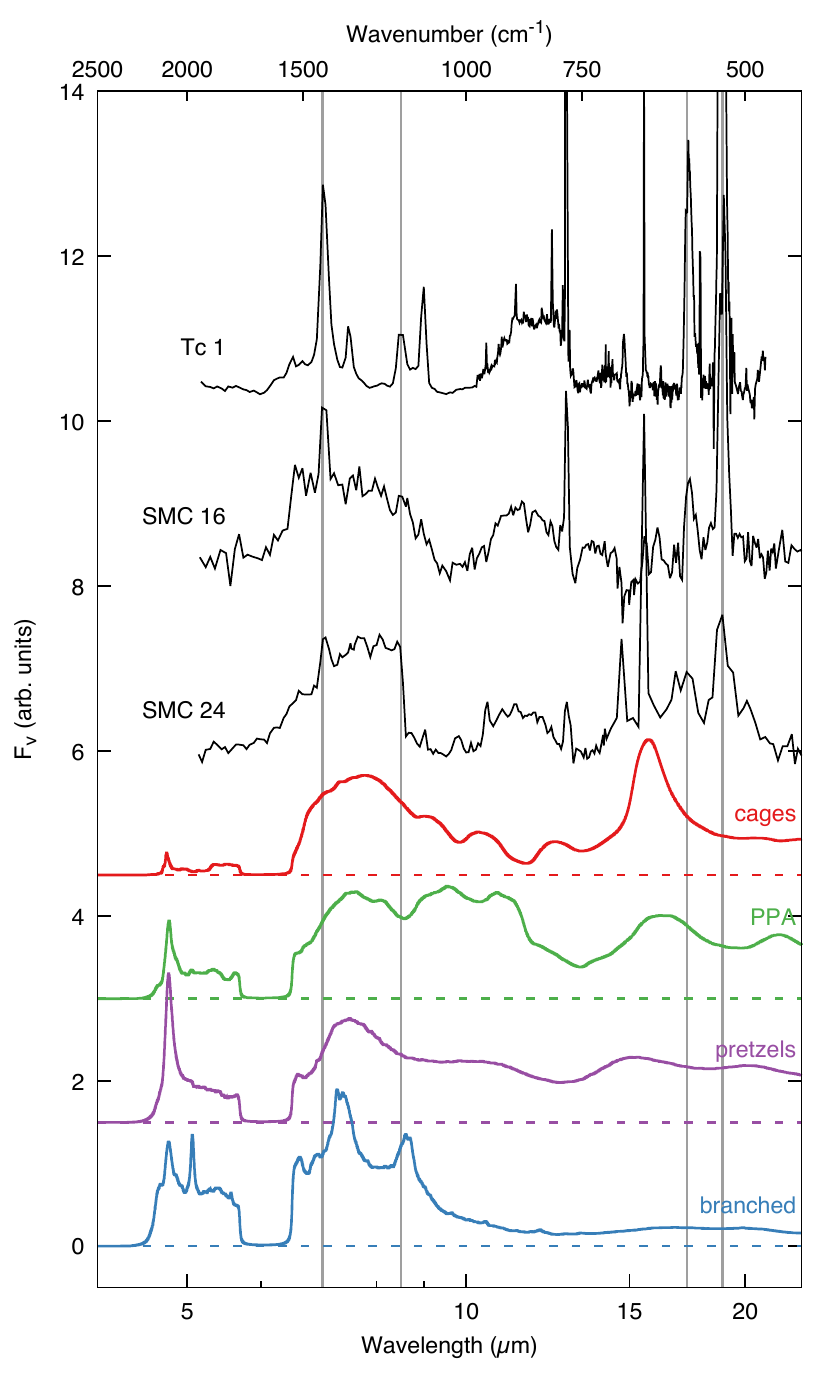}
   \caption{{\em Spitzer} astronomical data observed for the Tc 1 and SMP SMC 16 \citep{Hernandez2010} and SMP SMC 24 \citep{Hernandez2011b} planetary nebulae (black curves) and IR spectra calculated for the cages, PPAs, pretzels, and branched structures families (red, green, purple, and blue curves, respectively). The four vertical lines mark the IR active lines of buckminsterfullerene \citep{C60_vibexp_nature90}\label{fig:astro1}}
   \end{figure}

In the absence of clear information about the underlying thermodynamics under PNe and PPNe conditions, uniform populations were chosen to reconstruct the typical IR spectra of the four structural families from the individual contributions of each member. Such uniform distributions are an oversimplification of the energetics, although Fig.~\ref{fig1} indicates that cages, PPAs, and branched structures poorly overlap with each other in this respect. 
By summing over thousands of individual topologically distinct conformers, the spectra are necessarily broad and cannot account for sharp bands, which originate from very abundant specific structures and that must be assigned using more traditional approaches.  For comparison with astronomical data, the IR absorption intensities were further convoluted by a Planck blackbody radiation law at 400~K, as proposed earlier by Draine and Li \citep{Draine2001}. The four spectra thus determined for the four families are shown in the bottom part of Fig.~\ref{fig:astro1}.
        
All families exhibit features below 6~\si{\micro\meter}, although significantly attenuated by the convolution with the Planck law for the cages population. These lines involve sp$^1$ carbons found in surface (cages) or edge (PPA) defects, and long loops and linear chains (pretzels and branched). The  terminating chains, in particular, display a strong absorption near 4.7~\si{\micro\meter}. While such isomers exist among the cage [see, e.g., Fig.~\ref{fig1}(1)] and PPA [see, e.g., Fig.~\ref{fig1}(7)] populations, they are far more numerous in more open structures such as the pretzel and  the branched families, for which this peak is even more intense and doubled with another feature at 5.1~\si{\micro\meter}.

Above 6~\si{\micro\meter} the spectral features are due to the CC stretching of sp$^2$ carbons involved in multiple 5-, 6-, or 7-member rings. The corresponding active spectral region is broad and overlaps with the range where CCC in-plane bending modes are found near 10.5--12.5~\si{\micro\meter}. These features are most prominent for cages, for which the averaged spectrum exhibits a broad band between 6.2 and 11.6~\si{\micro\meter} with a sharp blue tail, a smooth red tail, and a sharper maximum near 6.6--7.9~\si{\micro\meter}. Smaller peaks near 8.9 and 10.3~\si{\micro\meter} are caused by CC stretchings, while the band near 12.5~\si{\micro\meter} is assigned to in-plane CCC bending modes leading to some rotation of the rings. Bands located above 14~\si{\micro\meter} correspond to softer deformations of the carbon skeleton with a more collective character. The accuracy of these lines is  disputable at the present level of theory (see Appendix A.2, Table A.1), and the narrow intense band appearing near 15.7~\si{\micro\meter} should actually be shifted by +1.2~\si{\micro\meter}.

The IR spectrum obtained for the PPA family shows a very broad band extending from 6.4 to 13.3~\si{\micro\meter}, with maxima at 7.4, 9.5, and 10.8~\si{\micro\meter}. These features are the signature of CC stretching modes that also increasingly combine with CCC bending modes as the wavelength increases. Two bands near 16.4 and 21.9~\si{\micro\meter} are assigned to in-plane and out-of-plane deformation modes, respectively.

Compared with cages and PPA isomers, the spectrum of the pretzel isomers above 6~\si{\micro\meter} displays fewer bands except broad maxima near 7.3 and 15.3~\si{\micro\meter} that we assign to CC stretchings and in-plane bendings, respectively. The much greater structural diversity among this population explains why these modes are associated with broader bands.

Finally, the spectrum of the branched family exhibits an active region in the 6.4--9.6~\si{\micro\meter} range, but hardly any feature above this range. The maxima at  7.3 and 8.6~\si{\micro\meter} are assigned to CC stretches coupled to in-plane CCC bending modes involving sp$^2$ carbons that pertain both to small rings and long loops or linear chains.

These spectra can now be compared to  astronomical observations recorded in  C-rich and H-rich PNes of the Small Magellanic Cloud (SMC), where fullerenes and PAHs have both been detected \citep{Hernandez2010} and where plateaus in the spectra have remained unassigned.
The emission features in SMP SMC 16 found in the 6-9~\si{\micro\meter} and 15-20~\si{\micro\meter} ranges  possibly match those found for the cage population, given that the band calculated at 15.7 \si{\micro\meter} should be shifted by about +1.2 \si{\micro\meter}. However, the contribution of cages to the 10-13 \si{\micro\meter} plateau is  questionable, as is the presence of PPA structures, whose spectral feature near 6-10~\si{\micro\meter} is much broader than the observed data, with no remarkable feature in the 10-13 \si{\micro\meter} range either. Similarly, pretzel isomers cannot explain the astronomical data for this PNe. In contrast, we find some similarities in the spectrum of the branched family near 6-9 \si{\micro\meter}, and especially its peak at 8.6~\si{\micro\meter}. 
However, the branched family spectrum shows no feature above 10-13  \si{\micro\meter}, in contradiction with observations.

Comparing now the calculated spectra with the astronomical observations of the Tc 1 PNe where \cbn{60} and \cbn{70} were detected \citep{cami2010}, the most stable cage population may also contribute to the same regions as for SMP-SMC 16, but cannot account for the very intense 10--13 \si{\micro\meter} plateau. The contribution of the other families must probably be excluded as they are too dissimilar with astronomical data for this PNe.

Comparison can also be made for the spectrum of SMP SMC 24, which was suggested as being due to a complex mixture of aliphatic and aromatic compounds such as HACs, PAH clusters, fullerenes, and small dehydrogenated carbon clusters \citep{Hernandez2011b}. As was the case for SMP-SMC 16, only the cage family is compatible as a potential contributor to the features near 6--9 \si{\micro\meter} and 15--20~\si{\micro\meter}. The spectra obtained for the cationic case of C$_{60}^+$ show essentially no change relative to the neutral system (see Appendix B, Fig.~\ref{figB}).  None of the studied families can account for the 10--13~\si{\micro\meter} feature. 
 However, additional calculations on smaller carbon clusters C$_{24}$ and C$_{42}$ show that their corresponding cage families possess more intense features in the 10-15 ~\si{\micro\meter} domain than C$_{60}$, relative to the 6-9~\si{\micro\meter} region (see Appendix C, Fig.~\ref{figC} top left panel). Two bands are observed in the spectrum of C$_{24}$ in the 10-13 ~\si{\micro\meter} range (11.1 ~\si{\micro\meter} and 11.8~\si{\micro\meter}) plus a weaker feature near 12.7-13.3 ~\si{\micro\meter}. The spectrum of C$_{42}$ presents a weak band   in the 10.7-12.2 ~\si{\micro\meter} range  plus a wide intense band centered at 13.7 ~\si{\micro\meter}.  These smaller cages could then also be considered as potential contributors to the 10--15~\si{\micro\meter} astronomical features. The  C$_{24}$ size is all the more interesting as it possesses features  in the 15--20~\si{\micro\meter} range (15.8, 16.8, 17.6, and 18.6 ~\si{\micro\meter} at our level of calculation) that are observed in SMC 16 and 24.  The present theoretical study providing the IR signature of  cage populations for 24, 42, and 60 carbon atoms thus complements recent density-functional-theory (DFT) investigations suggesting the possible individual contribution of closed cage fullerenes smaller than C$_{60}$ to the IR spectra of fullerene-PNe  \citep{Adjizian2016, Candian2019}.

Finally, emission spectra of  proto-planetary nebulae that represent the short evolutionary phase between AGB stars and PNe are also worth considering in our comparison, and here we take the example of the PPN IRAS
22272+5435  whose spectrum can be found in Fig. 3 of \cite{Kwok2001}. Interestingly, no intense band is observed at 4.7 \si{\micro\meter}, which discards compounds with long carbon chains such as pretzels or members of the branched family. The features near 8 and 12 \si{\micro\meter} that were assigned to the vibrations of alkanes or alkyle side groups pertaining to large carbonaceous particles \citep{Kwok2001} have no direct counterpart on our spectra. However, the shape of the 5-8 \si{\micro\meter} feature reported by \cite{Kwok2001} is consistent with the spectrum obtained here for the cages population. Moreover, other features such as the intense band near 17 \si{\micro\meter} are not observed in PPNe. As was the case for PNe, it is plausible that for such PPNe the hydrogenation state plays an important role in these IR features.

\section{Conclusions} \label{con}

Fullerenes identified in space necessarily originate from complex chemical rearrangements that proceed either by growing smaller scale species, or by reducing larger compounds until these aromatic cages with enhanced  stability are formed.
To unravel the possible intermediates in these rearrangements through their vibrational IR response, and without assuming any specific formation mechanism, we have attempted to systematically sample the possible isomers that a connected set of 60 carbon atoms can exhibit. The conformational landscape could be efficiently explored by combining REMD simulations employing the REBO potential, followed by systematic reoptimization and IR characterization using the more accurate SCC-DFTB method.

Different energetic, structural and electronic parameters allowed the $3\times 10^5$ structures obtained to be partitioned into four main families called  cages, planar polycyclic aromatics, pretzels, and branched. The IR spectra of these families exhibit significant differences from each other, which in turn could be exploited to get insight into their potential presence in planetary and proto-planetary nebulae. Comparison between calculated and astronomical spectra generally indicates that neutral or cationic cage isomers could account for the  6--9 \si{\micro\meter} plateau, possibly suggestive of defective fullerenes in these environments. Smaller cage  carbon clusters could also contribute. 
In contrast, more disordered structures containing long carbon loops or a greater number of linear chains are less likely, all the more as the intense IR feature found for these pretzel and branched families near 4.7~\si{\micro\meter} 
has not been observed in PNe and PPNe.

The main limitation of our  model is that only a single cluster size and stoichiometry were considered, without thermodynamical consideration within the isomer populations. Such important extensions are relatively straightforward to carry out using the present methodology, provided that the potential energy surfaces and IR spectra can be determined at a similar, reasonable computational cost. However, the large amount of associated data might require more robust analysis tools. In future developments, the contributions of hydrogen or other heteroatoms, different sizes and charge states, as well as anharmonicities, will be explored in order to help identify the structural patterns responsible for the as-yet-unexplained astronomical spectral features of PNe, notably in the 9-13 \si{\micro\meter} region. This work and its future developments will be of great importance to interpreting the IR spectra of carbon-rich PPNe and PNe soon to be observed by the James Webb Space Telescope.

\begin{acknowledgements}
The authors gratefully acknowledge financial support by the Agence Nationale de la Recherche (ANR) Grant No. ANR-16-CE29-0025, and the GDR EMIE 3533. The authors also acknowledge the computing mesocenter CALMIP (UMS CNRS 3667) for the generous allocation of computer resources (p0059) and Dr. D.A. Garc\'{\i}a Hern\'andez for discussions related to the astrophysical relevance of this work.
\end{acknowledgements}

\bibliographystyle{aa}
\bibliography{bib_carbo}

\newpage
\appendix

\section{Benchmark of IR spectra and correction factors}
\subsection{Systematic determination of correction factors}

The vibrational modes of carbon clusters are poorly known, except for a few exceptions such as buckminsterfullerene. Here we use density-functional theory with the B3LYP functional and the pc-1 basis set \citep{PC1} as the reference method. Given that this approach is not exact, all frequencies obtained with it are scaled by the appropriate factor of 0.973 \citep{Laury2012}.

\begin{figure}[H]
        \centering
        \includegraphics[scale=0.7]{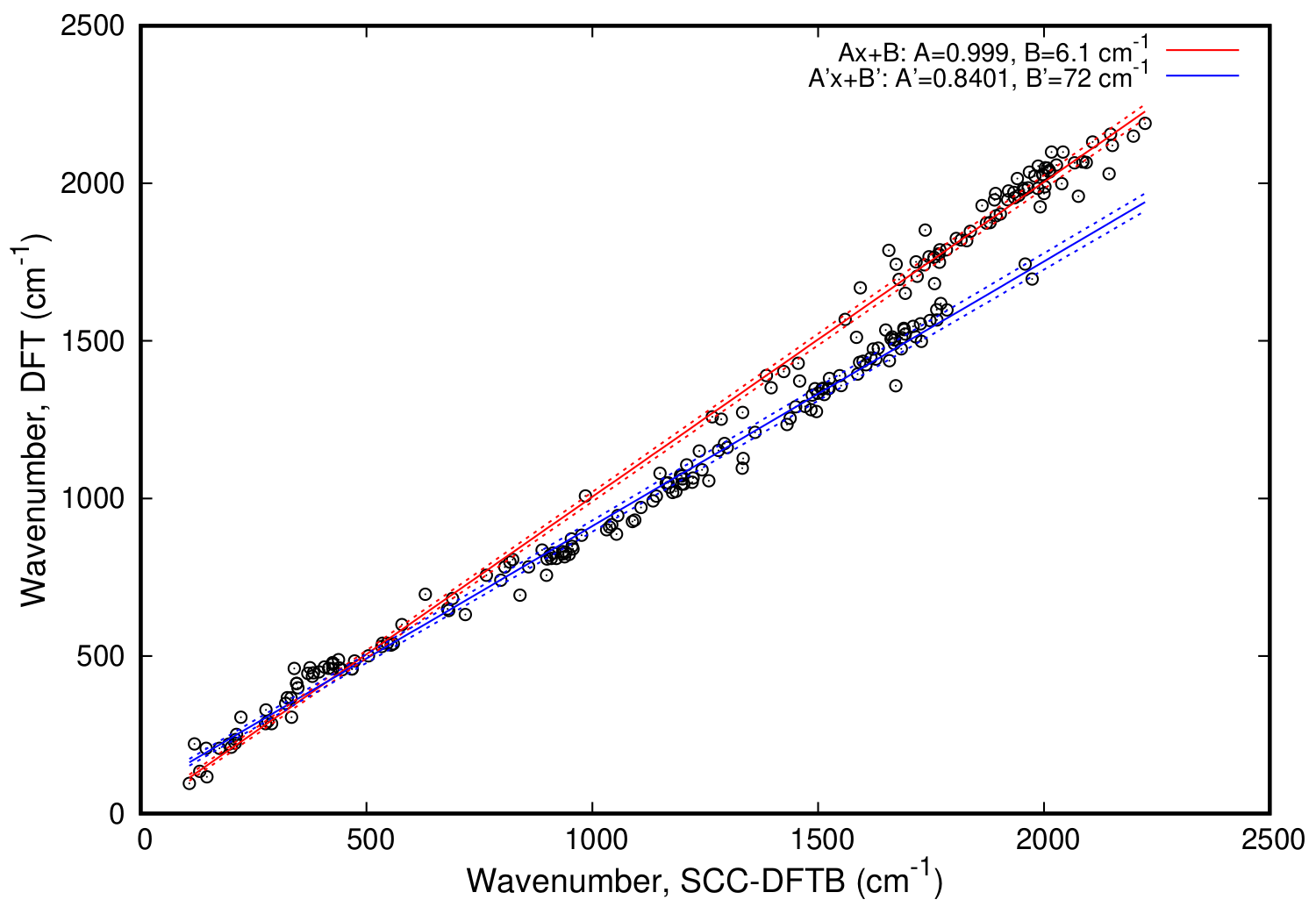}
    \caption{ Harmonic line positions obtained at the DFT/B3LYP/pc-1 level of theory against the corresponding values with the SCC-DFTB method, for a sample of IR active modes from 51 isomers. The dotted lines above and below the main lines illustrate the corresponding uncertainties on each of the linear least-squares fits.}
    \label{figA1}
\end{figure}

 A  sample of 50 isomers of \cbn{24} were chosen randomly among those reported in the work by~\citet{bonnin2019}. 
 Each isomer was locally reoptimized at the DFT/B3LYP/pc-1 level of theory, the normal modes being determined by standard diagonalization of the mass-weighted Hessian matrix. DFT  calculations were performed with the Gaussian09 suite of programs \citep{g09}. This was not achieved for \cbn{60} isomers due to the  too high computational cost at the DFT level. However, the  peculiar but well known buckminsterfullerene isomer of \cbn{60} was added to the sample. Correlations between the frequencies obtained with the two methods (DFTB and DFT) can be seen for a subset of particularly active modes in Fig.~\ref{figA1}. 

This correlation plot displays two main linear regions, depending on whether the active mode lies below 1750~\si{\per\centi\meter} or above this approximate threshold. The DFT frequencies are recovered from SCC-DFTB frequencies if linear relationships of the type 
$$\sigma_{\rm DFT} = A\sigma_{\rm DFTB}+B$$ 
are used in the corresponding ranges.

Least-squares fitting procedures readily provide the following values for the linear coefficients, namely
$$A=0.8401\pm 0.0080 \mbox{~~and~~} B=72\pm 10~\si{\per\centi\meter}$$
for $\sigma_{\rm DFT}<1750$~\si{\per\centi\meter}, and
$$A=0.9990\pm 0.0061 \mbox{~~and~~} B=6.1\pm 9.3~\si{\per\centi\meter}$$
for $\sigma_{\rm DFT}>1750$~\si{\per\centi\meter}.

From the knowledge of the DFTB values, the IR band positions used for comparison with astronomical spectra were determined using these linear relations, and further scaled by 0.973. For the sake of clarity, the uncertainty on the determined correction factors also shown in Fig.~\ref{figA1} was not applied to the spectra of Fig. 2.

\subsection{Assessment of linear corrections on buckminsterfullerene and fully dehydrogenated coronene}

The IR active modes of icosahedral \cbn{60} (buckminsterfullerene) and $D_{6h}$ fully dehydrogenated coronene \cbn{24} were determined with the SCC-DFTB and DFT/B3LYP/pc-1 methods. The raw positions directly obtained with these methods are listed in Tables A.1 and A.2, along with the corrected SCC-DFTB values. For buckminsterfullerene, the residual shifts between corrected SCC-DFTB values and experimental data are also given.

\begin{table}[H]
        \centering
        \begin{tabular}{cccccc}
        \hline\hline
Exp. & Raw & Raw & Corrected & Residual \\
 & B3LYP/pc-1 & SCC-DFTB & SCC-DFTB & shift to exp.\\
        \hline
18.9  & 18.7 & 19.8 & 20.7 & +1.8 \\ 
17.3  & 17.0 & 14.9 & 16.1 & -1.2 \\
 8.5  & 8.3 & 7.1 & 8.3 & -0.2 \\
 7.0  & 6.9 & 6.0 & 7.0 & 0.0 \\
\hline
\end{tabular}
\caption{Raw resonance wavelengths (in \si{\micro\meter}) of C$_{60}$ buckminsterfullerene IR active normal modes at the DFT and SCC-DFTB levels. The corrected SCC-DFTB values are also mentioned, along with the corresponding shifts with respect to experimental data \citep{C60_vibexp_nature90}.}
\end{table}

\begin{table}[H]
        \centering
\begin{tabular}{ccc}
        \hline\hline
Raw & Raw & Corrected \\ 
B3LYP/pc-1 & SCC-DFTB & SCC-DFTB  \\ 
        \hline
24.2  & 29.0 & 28.4  \\ 
21.8  & 21.4 & 22.1  \\ 
12.4  & 11.0 & 12.3  \\ 
9.5   & 8.2 & 9.4\\ 
6.5   & 5.8 & 6.8\\ 
        \hline
\end{tabular}
        \caption{Raw resonance wavelengths (in \si{\micro\meter}) of dehydrogenated coronene C$_{24}$ IR active normal modes at the DFT and SCC-DFTB levels. The corrected SCC-DFTB values are also mentioned.}
\end{table}

\section{ Structural isomers of C$_{60}$$^+$: IR spectra}

\begin{figure}[H]
        \centering
        \includegraphics[scale=0.7]{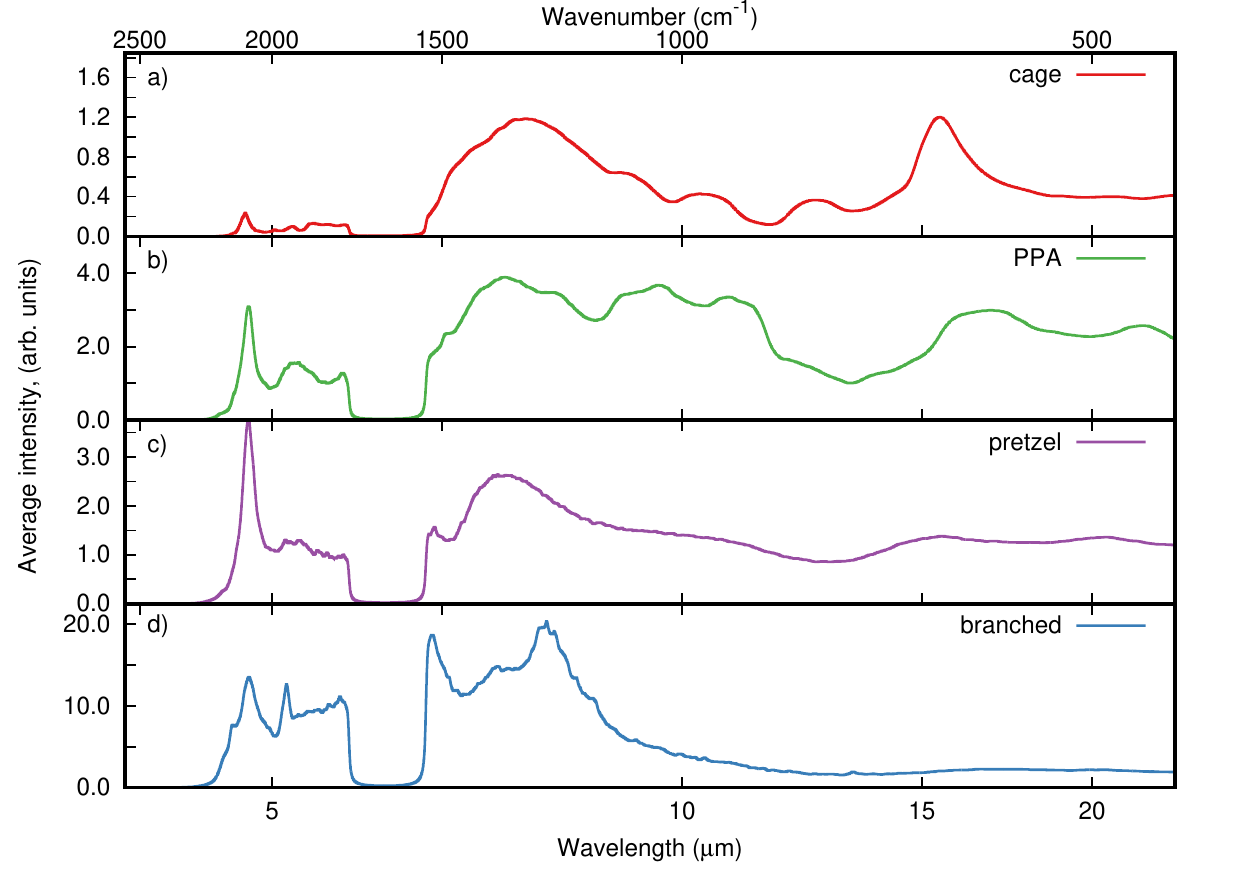}
    \caption{ SCC-DFTB IR spectra of the families determined for C$_{60}$$^+$ }
    \label{figB}
    \end{figure}

The geometries of all structural isomers of C$_{60}$ obtained at the SCC-DFTB level were optimized in a positive charge state (doublet spin-state). Only slight  structural changes were observed, the positive charge being  diluted over the 60 carbon atoms. Their IR spectra were computed in the double harmonic approximation similarly to their neutral counterparts and the resulting average spectra for the four families as defined in the main text are shown in Fig. \ref{figB}.  
    
    \section{Smaller size carbon clusters (C$_{42}$ and C$_{24}$): IR spectra }
    
\begin{figure*}
        \centering
        \includegraphics[scale=1.5]{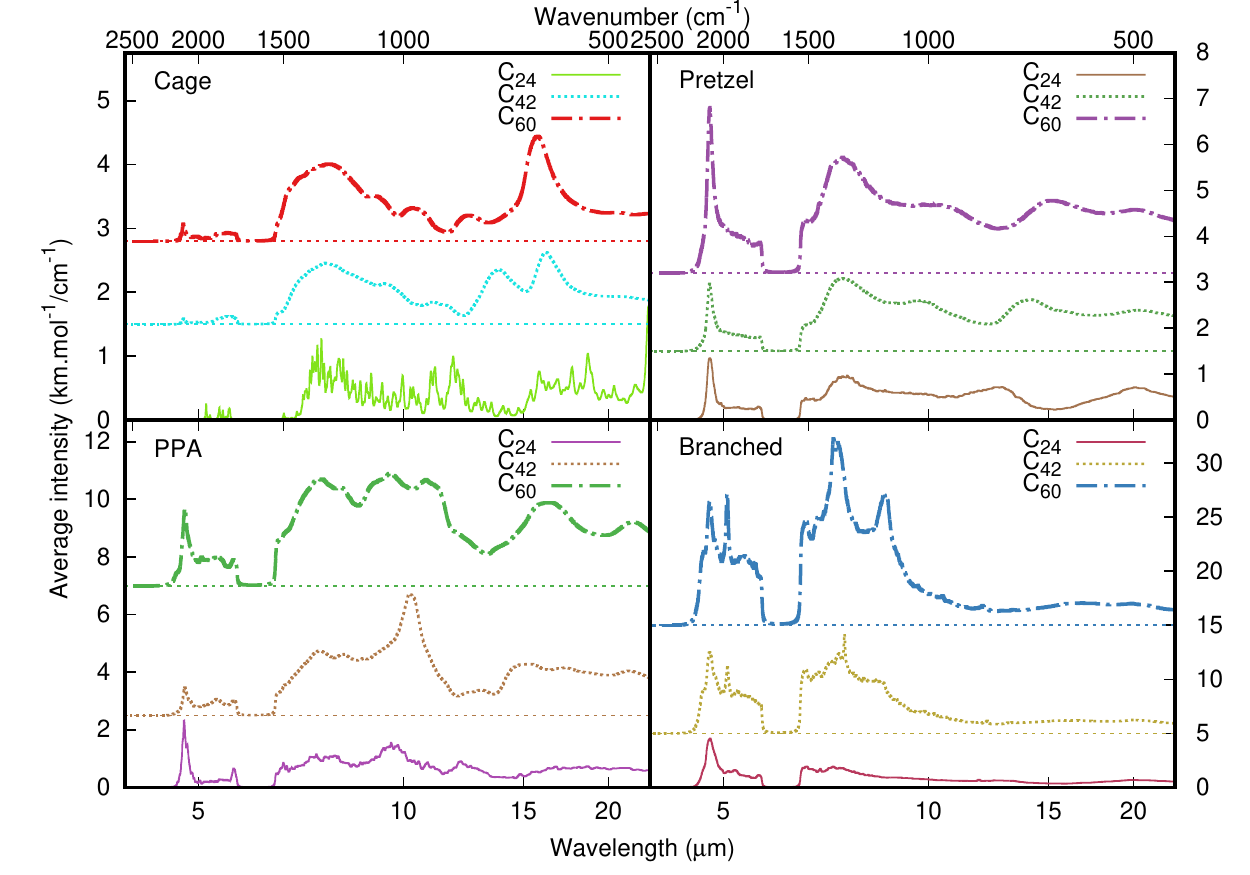}
    \caption{SCC-DFTB IR spectra of the isomer families of C$_{24}$ and C$_{42}$. Number of isomers per family, C$_{42}$: 27324 cages, 40914 PPAs, 36036 pretzels, and 94130 branched structures; C$_{24}$: 11 cages, 714 PPAs, 6307 pretzels, and 37309 branched structures.}
    \label{figC}
\end{figure*}

    A similar study to that proposed in the present letter for C$_{60}$ has been performed for smaller clusters, {i.e.,} C$_{42}$ and C$_{24}$ with initial REBO structures taken from ~\citet{bonnin2019}. The criteria used to determine the families are similar to those described for C$_{60}$. The IR spectra of the four families for the three cluster sizes are given in \ref{figC}.

\end{document}